\documentclass[
superscriptaddress,
nofootinbib,
twocolumn,
 amsmath,amssymb,
 aps,
pra,
notitlepage,
longbibliography
]{revtex4-1}

\usepackage{dcolumn}
\usepackage{bm}
\usepackage{epsfig}
\usepackage{graphicx}
\usepackage{latexsym}
\usepackage{amsfonts}
\usepackage{float} 
\usepackage{setspace}
\usepackage{graphicx}
\usepackage{amsmath}
\usepackage{mathrsfs}
\usepackage{verbatim}
\usepackage{color}
\usepackage{SIunits}
\usepackage{hyperref}

\usepackage{physics}
\usepackage{soul}

\newcommand{\be}{\begin{equation}}
\newcommand{\ee}{\end{equation}}
\newcommand{\ba}{\begin{array}}
\newcommand{\ea}{\end{array}}
\newcommand{\bqa}{\begin{eqnarray}}
\newcommand{\eqa}{\end{eqnarray}}

\renewcommand{\arraystretch}{1.2}

\begin{document}

\title{Quantum-coherent optical isolation and circulation using frequency conversion on a chip}

\author{Jierui Hu} 
\thanks{These authors contributed equally to this work.}
\affiliation{Holonyak Micro and Nanotechnology Laboratory and Department of Electrical and Computer Engineering, University of Illinois at Urbana-Champaign, Urbana, IL 61801 USA}
\affiliation{Illinois Quantum Information Science and Technology Center, University of Illinois at Urbana-Champaign, Urbana, IL 61801 USA}
\author{Hao Yuan} 
\thanks{These authors contributed equally to this work.}
\affiliation{Holonyak Micro and Nanotechnology Laboratory and Department of Electrical and Computer Engineering, University of Illinois at Urbana-Champaign, Urbana, IL 61801 USA}
\affiliation{Illinois Quantum Information Science and Technology Center, University of Illinois at Urbana-Champaign, Urbana, IL 61801 USA}
\author{Joshua Akin} 
\thanks{These authors contributed equally to this work.}
\affiliation{Holonyak Micro and Nanotechnology Laboratory and Department of Electrical and Computer Engineering, University of Illinois at Urbana-Champaign, Urbana, IL 61801 USA}
\affiliation{Illinois Quantum Information Science and Technology Center, University of Illinois at Urbana-Champaign, Urbana, IL 61801 USA}
\author{Shanhui Fan} 
\affiliation{Department of Electrical Engineering, Ginzton Laboratory, Stanford University, Stanford, CA 94305, USA}
\author{Kejie Fang} 
\email{kfang3@illinois.edu}
\affiliation{Holonyak Micro and Nanotechnology Laboratory and Department of Electrical and Computer Engineering, University of Illinois at Urbana-Champaign, Urbana, IL 61801 USA}
\affiliation{Illinois Quantum Information Science and Technology Center, University of Illinois at Urbana-Champaign, Urbana, IL 61801 USA}

\begin{abstract} 

Breaking optical reciprocity enables new regimes of light--matter interaction with broad implications for fundamental physics and emerging quantum technologies. Although various approaches have been explored to achieve optical nonreciprocity, realizing it at the single-photon level has remained a major challenge. Here, we demonstrate nonmagnetic optical nonreciprocity---including both isolation and circulation---in the quantum regime, enabled by efficient and noiseless all-optical frequency conversion on an integrated III-V  photonic chip. Our device preserves the quantum coherence and entanglement of the input photons while delivering exceptional performance parameters, including a high extinction ratio of 34 dB, low insertion loss of 0.8 dB, broad bandwidth of 44 GHz, high operational fidelity of 97\%, and widely tunable operation wavelength. This realization of quantum optical nonreciprocity in a scalable photonic platform opens a pathway toward directional quantum communication and noise-resilient quantum networks. 

\end{abstract}

\maketitle

Reciprocity, a cornerstone of Maxwell's equations, ensures symmetric light transmission and forms the basis of most conventional optical systems. Breaking this fundamental symmetry of electromagnetic reciprocity reveals new physical phenomena \cite{haldane2008possible,wang2009observation,fang2012realizing,lu2014topological} and enables novel functionalities in both classical and quantum regimes. For example, once this symmetry is broken, it becomes possible to engineer nonreciprocal devices---components that allow light to transmit preferentially in one direction---such as optical isolators \cite{jalas2013and}, which block unwanted backscattering and are critical for stabilizing lasers and amplifiers in communication networks \cite{li2002impact}. In the quantum regime, the ability to control electromagnetic reciprocity becomes even more significant, as nonreciprocal components can protect superconducting qubits from amplifier noise and enable quantum-limited directional amplification in quantum microwave circuits \cite{kamal2011noiseless,metelmann2015nonreciprocal,kerckhoff2015chip,lecocq2017nonreciprocal,bernier2017nonreciprocal,ranzani2017wideband,navarathna2023passive}. In optical quantum systems, nonreciprocal devices are increasingly sought after  \cite{sayrin2015nanophotonic,scheucher2016quantum,dong2021all,hu2021noiseless,zhang2023noiseless,wang2025self,zhang2025chirality,ren2022single}, where they can protect sensitive quantum components from stray light in a network and enable new forms of light--matter interaction for quantum simulation, sensing, and communication \cite{stannigel2012driven,gonzalez2015chiral,lodahl2017chiral,huang2018nonreciprocal,barik2018topological,lau2018fundamental,mahmoodian2020dynamics,ahmadi2024nonreciprocal}.

A variety of approaches have been explored to achieve optical nonreciprocity, including magneto-optic effects \cite{bi2011chip,ren2022single} and passive optical nonlinearities \cite{fan2012all,white2023integrated}. However, magneto-optic materials are intrinsically lossy and difficult to integrate on-chip, while passive optical nonlinearities are too weak to produce nonreciprocity at the single-photon level. Recently, atomic systems have been shown to be effective for quantum optical nonreciprocity \cite{sayrin2015nanophotonic,scheucher2016quantum,dong2021all,hu2021noiseless,zhang2023noiseless,wang2025self,zhang2025chirality}, but they remain constrained by fixed operational wavelengths, narrow bandwidths, and limited scalability. Another widely investigated strategy for inducing optical nonreciprocity is spatiotemporal modulation \cite{yu2009complete,tzuang2014non,shen2016experimental,fang2017generalized,kittlaus2018non,sohn2021electrically,tian2021magnetic,kittlaus2021electrically,cheng2025terahertz,shah2023visible,herrmann2022mirror,yu2023integrated,sounas2017non}.
Unlike passive approaches, spatiotemporal modulation requires external driving fields to break time-reversal symmetry. Whether such actively driven nonreciprocal optical systems can function in the quantum regime---where stringent constraints on loss and noise apply---remains an open question.

Here, we demonstrate the first nonmagnetic optical nonreciprocity---encompassing both isolation and circulation---at the single-photon level on an integrated photonic chip, while preserving the quantum coherence and entanglement of the input photons. Our device employs optical frequency conversion in nonlinear nanophotonic waveguides with parametric optical pumps, achieving an added noise far below the single-photon limit. Thanks to the efficient optical frequency conversion, the pump power required for high isolation is significantly lower than that of spatiotemporally modulated isolators using electro-optic or acousto-optic effects \cite{cheng2025terahertz,tian2021magnetic,sohn2021electrically,kittlaus2021electrically,kittlaus2018non,shah2023visible,yu2023integrated,herrmann2022mirror}.   Our platform simultaneously achieves a high extinction ratio of 34 dB, low insertion loss of 0.8 dB, broad bandwidth of 44 GHz, high operational fidelity of 97\% (for the circulator), and widely tunable operation wavelengths, significantly outperforming previous demonstrations of quantum optical nonreciprocity based on atomic systems  \cite{sayrin2015nanophotonic,scheucher2016quantum,dong2021all,hu2021noiseless,zhang2023noiseless,wang2025self,zhang2025chirality} and magneto-optic effects \cite{ren2022single}. These results demonstrate the feasibility of spatiotemporally modulated optical nonreciprocity in the quantum regime and advance the path toward robust, integrated quantum optical systems with broken time-reversal symmetry. Our approach based on parametrically driven optical nonlinearity also provides a solution for integrated isolators on III-V semiconductor platforms, which are constrained by weak electro-optic and piezoelectric effects.

\begin{figure*}[!htb]
	\begin{center}
		\includegraphics[width=2\columnwidth]{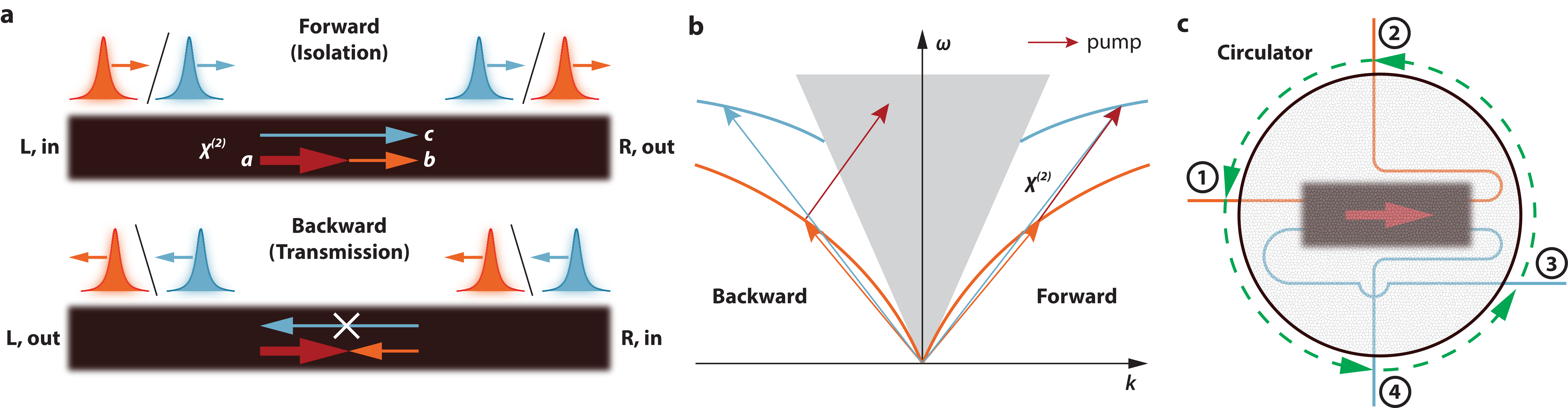}
		\caption{\textbf{Optical nonreciprocity via noiseless optical frequency conversion.} 
			\textbf{a}. Illustration of optical isolation via optical frequency conversion in a $\chi^{(2)}$ nonlinear waveguide. \textbf{b}. Representation of $\chi^{(2)}$-mediated optical frequency conversion in the band diagram. Only the forward direction satisfies the frequency- and phase-matching conditions for the frequency conversion.  \textbf{c}. Optical circulation utilizing the full mode space of a parametrically-pumped $\chi^{(2)}$ waveguide. \textbf{d}. InGaP photonic chip. \textbf{e}. 3D illustration of the phase-matching tunable waveguide device (not to scale).  \textbf{f}. Optical microscope image of an InGaP nanophotonic waveguide integrated with a nanoheater array. \textbf{g}. Scanning electron microscope image of a portion of the device and simulated field profile of the waveguide modes.   }
		\label{fig1}
	\end{center}
\end{figure*}

\noindent\textbf{Concept}\\
Our quantum optical isolator and circulator are realized on the basis of noiseless parametric frequency conversion \cite{kumar1990quantum} in a waveguide with $\chi^{(2)}$ nonlinearity (Figs. \ref{fig1}a and b), although the principle can also be extended to systems with $\chi^{(3)}$ nonlinearity. Consider a waveguide supporting three optical modes, $a$, $b$, and $c$, where the $\chi^{(2)}$ nonlinearity facilitates a three-wave mixing process described by the Hamiltonian
$\hat H = \hbar \chi(\hat a^\dagger \hat b^\dagger \hat c + \hat a \hat b \hat c^\dagger)$
with $\hat a$ ($\hat a^\dagger$) denoting the annihilation (creation) operator for mode $a$, and similarly for modes $b$ and $c$.
In the forward direction, when the three modes satisfy the frequency- and phase-matching conditions $\omega_a + \omega_b = \omega_c$ and $k_a + k_b = k_c$, where $\omega$ and $k$ denote the angular frequency and wavevector, respectively, and mode $a$ is driven by a  classical pump field with amplitude $\alpha$, the interaction between modes $b$ and $c$ becomes linearized, i.e., $\hat H = \hbar \chi\alpha(\hat b^\dagger \hat c +  \hat b \hat c^\dagger)$, which enables noiseless frequency conversion between the two modes, i.e., sum-frequency generation (SFG) ($b\rightarrow c$) and difference-frequency generation ($c \rightarrow b$). 

\begin{figure*}[!htb]
	\begin{center}
		\includegraphics[width=2\columnwidth]{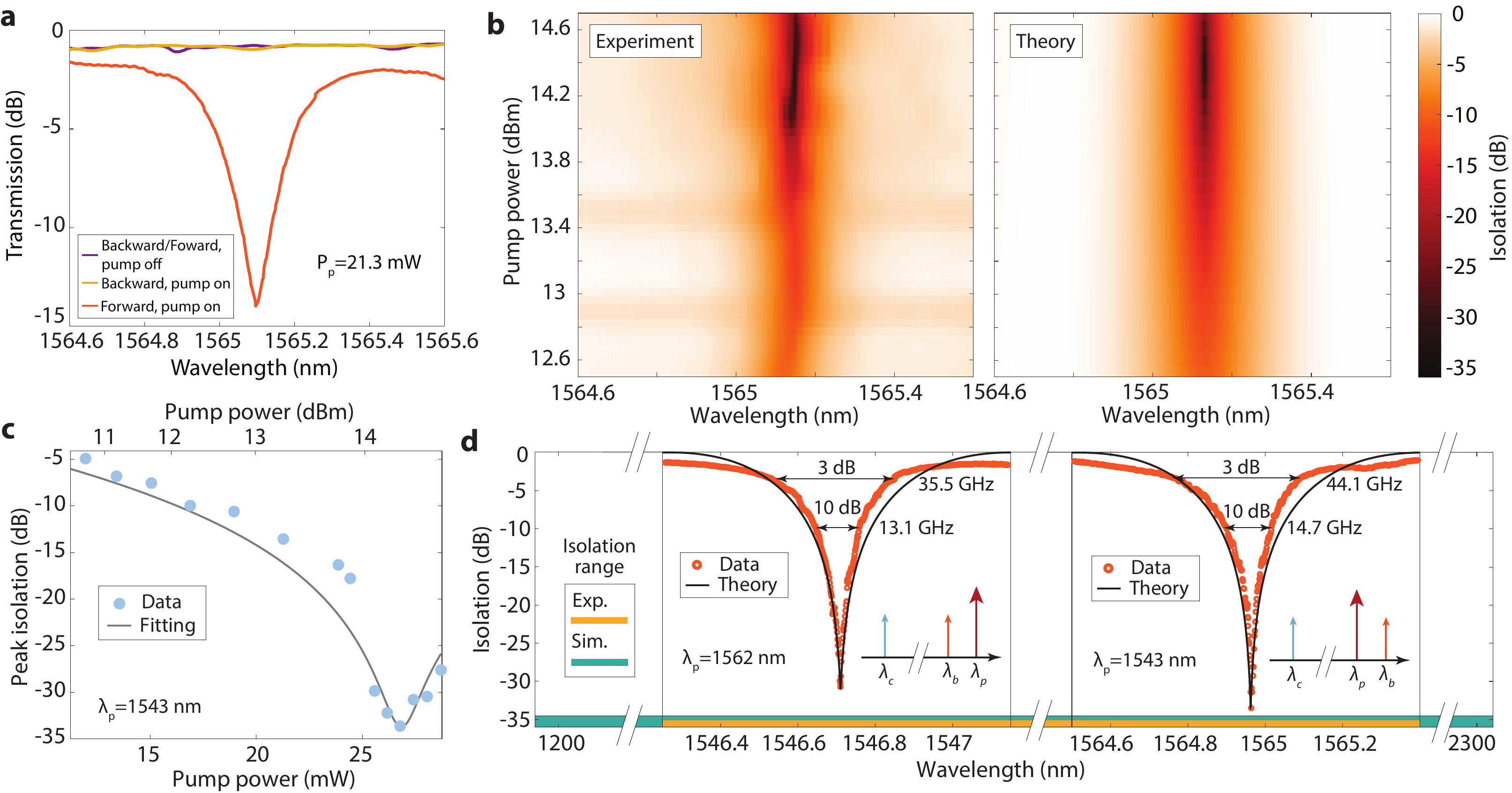}
		\caption{\textbf{Isolator performance.} 
			\textbf{a}. Backward and forward transmissions of the 6-mm isolator for pump on and off with pump power of 21.3 mW.  \textbf{b}. Measured and theoretical isolation versus pump power. \textbf{c}. Peak isolation ratio versus pump power.  \textbf{d}. Tunable isolation by changing the pump wavelength. Two isolation spectra are shown for pump wavelength of 1562 nm and 1543 nm, respectively. The 3-dB and 10-dB isolation bandwidths are indicated. Measured isolation wavelength range is 1538 nm$-$1570 nm due to the available laser wavelengths while the simulated isolation range is $> 1000$ nm for the same device. }
		\label{fig2}
	\end{center}
\end{figure*}

When the frequency conversion is efficient, the input photons in mode $b\ (c)$ can be completely transferred to mode $c\ (b)$. Thus, filtering out the pump and light in mode $c\ (b)$ effectively blocks the forward transmission of photons in mode $b\ (c)$. In the backward direction, the same pump configuration no longer satisfies the phase-matching condition, and frequency conversion between modes $b$ and $c$ is suppressed. Consequently, the light input in mode $b\ (c)$ is transmitted, resulting in nonreciprocal transmission. It should be noted that in this context, the definitions of forward and backward directions are opposite to those typically used for an isolator. Importantly, this isolator operates simultaneously in widely separated optical frequency bands and inherently integrates quantum frequency conversion, making it particularly attractive for hybrid quantum networks involving photons of disparate wavelengths.  When the full mode space is considered, the scattering matrix of the device, defined by $(b_{\mathrm{L},\mathrm{out}},b_{\mathrm{R},\mathrm{out}},c_{\mathrm{L},\mathrm{out}},c_{\mathrm{R},\mathrm{out}})^T=\mathcal{S}(b_{\mathrm{L},\mathrm{in}},b_{\mathrm{R},\mathrm{in}},c_{\mathrm{L},\mathrm{in}},c_{\mathrm{R},\mathrm{in}})^T$, becomes  asymmetric---a key feature of optical nonreciprocity \cite{jalas2013and}:
\be\label{smatrix}
\setlength{\arraycolsep}{10pt}
\mathcal{S}=\begin{bmatrix}
 0 & 1 & 0  & 0 \\
0 & 0  & 1 & 0 \\
0 & 0 & 0 & 1 \\
1 & 0 &  0 &  0
\end{bmatrix}.
\ee
This scattering matrix indicates that the $\chi^{(2)}$ waveguide can also function as a four-port optical circulator when all modes are utilized (Fig. \ref{fig1}c), with $(b_{\mathrm{L}},b_{\mathrm{R}},c_{\mathrm{L}},c_{\mathrm{R}})$ corresponding to ports $1-4$, respectively. Furthermore, the directionality of both isolation and circulation can be reversed by switching the propagation direction of the pump.

This form of optical nonreciprocity, which is based on $\chi^{(2)}$- or $\chi^{(3)}$-mediated parametric frequency conversion, is distinct from passive nonlinear optical approaches \cite{fan2012all,white2023integrated}. Because the interaction is linearized by a classical pump, it circumvents the dynamic reciprocity constraint that limits passive nonlinear isolators \cite{shi2015limitations}. Moreover, unlike nonreciprocity via parametric down-conversion \cite{abdelsalam2020linear}---which relies on pumping the highest-frequency mode and thus suffers from spontaneous emission noise---our approach leverages noiseless conversion while preserving the original signal frequency in the transmission direction when operating as an isolator. These are critical for the realization of quantum optical isolation, as they avoid the need for single-photon nonlinearities and allow seamless downstream information processing.

\noindent\textbf{Integrated nonlinear photonics platform}\\
For a waveguide of length $L$, parametric frequency conversion causes Rabi oscillation between the two modes governed by the pump power. The isolation ratio is given by  (Supplementary Information (SI)) 
\be
\mathcal{I}=\frac{\Delta k^2}{4g^2+\Delta k^2}+\frac{4g^2}{4g^2+\Delta k^2}\cos^2\sqrt{g^2+\frac{\Delta k^2}{4}}L,
\ee
where $g=\sqrt{\frac{\omega_b}{\omega_c}\eta_{\mathrm{SFG}}P_a}$, $\eta_{\mathrm{SFG}}$ is the SFG efficiency in the weak pump limit, and $\Delta k=k_c-k_a - k_b$ is the wavelength-dependent phase mismatch between the optical modes. Maximum isolation occurs at half of the oscillation period for specific pump powers, where the energy transfer between the modes is optimized, with the maximum isolation ratio given by
\be
\mathcal{I}_{\mathrm{max}}=\left((2n+1)\frac{\pi}{L}\right)^2/\Delta k^2,
\ee
for integer $n\geq0$. The maximum isolation ratio is fundamentally constrained by the waveguide's phase mismatch and is realized when $n + \frac{1}{2}$ oscillations occur between modes $b$ and $c$.

\begin{figure*}[!htb]
	\begin{center}
		\includegraphics[width=2\columnwidth]{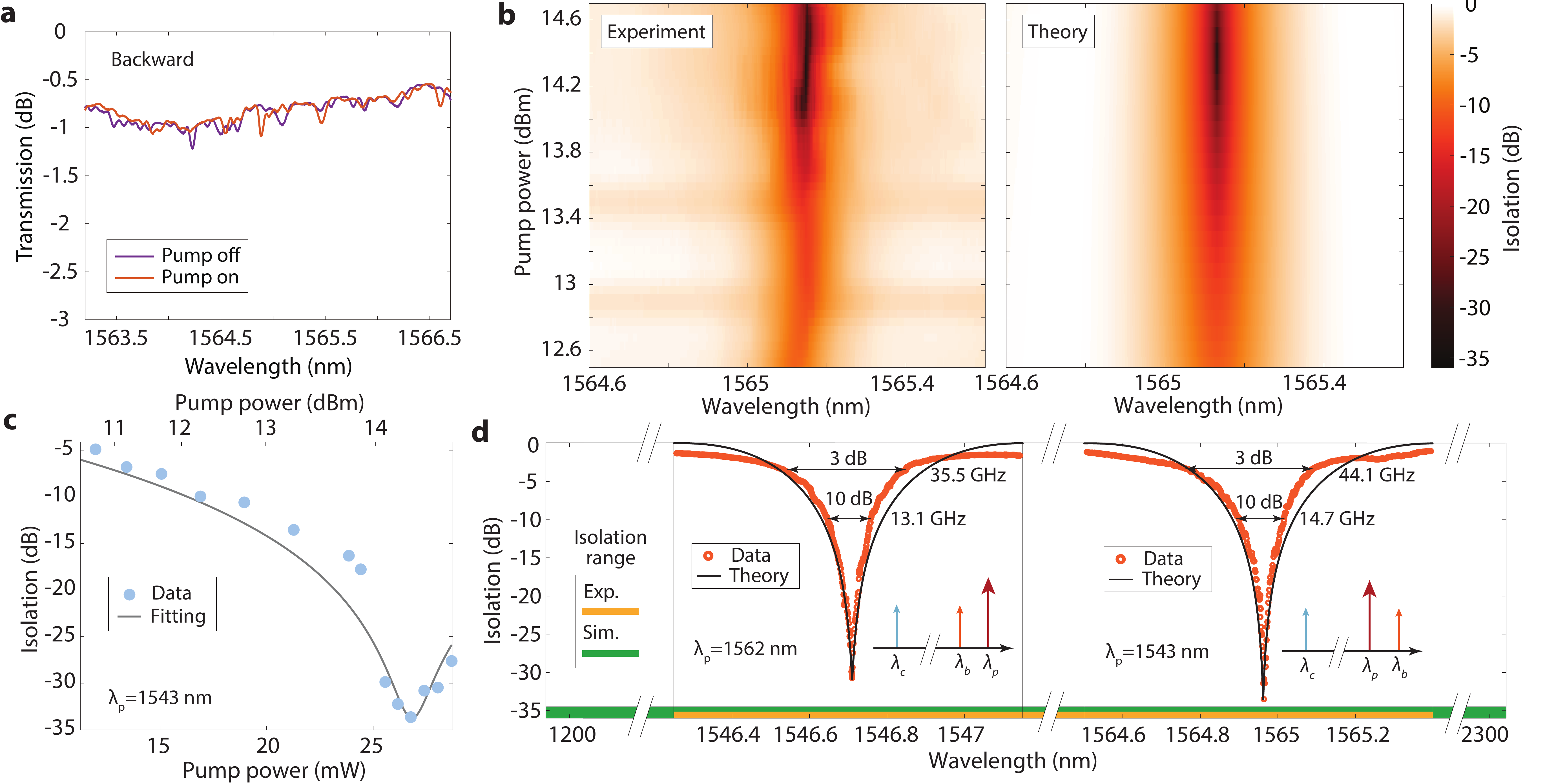}
		\caption{\textbf{Quantum optical isolator and circulator.} 
			\textbf{a}. Forward and backward on-chip noise flux in the 1550-nm band at the wavelength separated from the pump by 20 nm ($-2.5$ THz) versus the pump power of the 6-mm waveguide. The line is a linear fitting.  \textbf{b}. Normalized correlation function $g^{(2)}(\tau)$ of non-degenerate correlated photon pairs in the forward direction for pump on and off. EPS: entangled-photon source. Line is fitting. \textbf{c}. Two-photon interference fringe of degenerate correlated photon pairs after transmission through the optical isolator in the backward direction. Time-resolved coincidence counts are collected in 250-ps time-bins and for 180 s. The visibility of the fringe is $91.7\%$. \textbf{d}. Measured transmission matrix in the counterclockwise direction. The operational fidelity is $\mathcal{F}=0.962(4)$. \textbf{e}.  Measured transmission matrix in the clockwise direction. The operational fidelity is $\mathcal{F}=0.970(3)$.  \textbf{f}. Total and noise-subtracted coincidence counts in 2.3 GHz bandwidth in 60 s versus entangled photon pair rate after the signal is transmitted from port 1 to port 4 in the counterclockwise circulator. }
		\label{fig3}
	\end{center}
\end{figure*}

We demonstrate the optical nonreciprocity using the III-V InGaP integrated photonics platform (Figs. \ref{fig1}d-e) with a substantial $\chi^{(2)}$ nonlinearity ($\chi^{(2)}=220$ pm/V) and low optical losses \cite{zhao2022ingap, akin2024ingap,akin2024perspectives} (see the SI for device fabrication). The InGaP nanophotonic waveguide is designed to be modal-phase-matched between the 1550-nm-band fundamental transverse electric mode (for $a$ and $b$) and the 780-nm-band fundamental transverse magnetic mode (for $c$). For a phase-matched InGaP nanophotonic waveguide along the (110) direction, the simulated SFG efficiency at the weak pump limit is $\eta_{\mathrm{SFG}}\equiv\frac{P_c}{P_aP_bL^2} = 520,000\%$/W/cm$^2$. We also implemented in situ phase-matching tuning nanoheater arrays (Figs. \ref{fig1}f-g) to control the temperature of each segment of the waveguide and counteract the effect of thickness variations on the phase-matching condition (SI). By minimizing the phase mismatch along the entire waveguide, a high isolation ratio can be achieved.

\noindent\textbf{Isolator performance}\\
We use the phase-matching optimized 6-mm-long waveguide to demonstrate the optical isolator. The pump and signal are supplied by two wavelength-tunable telecom lasers, respectively. The pump and the converted light in the 780-nm band are filtered for the transmission measurement of the signal. Fig.~\ref{fig2}a shows the backward and forward transmissions when the pump is switched on and off, with the pump power of 21.3 mW. The backward transmission remains unchanged with an average insertion loss of 0.8 dB due to the waveguide propagation loss in the 1550-nm band. 
Fig.~\ref{fig2}b shows the measured isolation, $\mathcal{I}[\lambda]=|t_f[\lambda]/t_b[\lambda]|^2$, where $t_{f(b)}[\lambda]$ is the forward (backward) transmission coefficient as a function of wavelength, and the modeled isolation at approximately 1565 nm for various pump powers and pump wavelength at 1543 nm (see SI for the forward and backward  transmission spectra).  Fig. \ref{fig2}c shows the corresponding peak isolation ratio. The isolation ratio increases with pump power and reaches a maximum of 33.6 dB at a pump power of 27 mW, corresponding to the first half oscillation between the 1550-nm- and 780-nm-band modes. The maximum isolation is limited by the residual phase mismatch in the waveguide at length scales shorter than that of individual nanoheaters and can be improved with more discrete nanoheaters. The 3-dB and 10-dB isolation bandwidths corresponding to the maximum isolation are 44.1 GHz and 14.7 GHz, respectively. The signal power used in this measurement is approximately 17 $\mu$W, which is much weaker than that of the pump. We also measured isolation for various signal power in the range of 1 $\mu$W--200 $\mu$W and the maximum isolation is similar (SI).

The isolation wavelengths, $\lambda_b$ and $\lambda_c$, of the dual-band isolator can be tuned by adjusting the pump wavelength, $\lambda_p$, since the isolation is determined by the frequency- and phase-matching conditions of the frequency conversion: $\frac{1}{\lambda_p}+\frac{1}{\lambda_b}=\frac{1}{\lambda_c}$ and $\frac{n_p}{\lambda_p}+\frac{n_b}{\lambda_b}=\frac{n_c}{\lambda_c}$, where $n_k$ is the effective index of the respective mode and wavelength. Numerical simulations indicate that the fabricated waveguide supports wide isolation wavelength ranges, with mode $b$ tunable over more than 1000 nm and mode $c$ over more than 30 nm. Due to the available laser wavelengths in our experiment, we demonstrate the tuning of the isolation wavelength by setting the pump at 1562 nm and reoptimizing the phase-matching condition of the waveguide using the nanoheater array. Under these conditions, a maximum isolation of 30.3 dB is achieved at 1546.7 nm with a pump power of 29 mW, as shown in Fig. \ref{fig2}d.

\noindent\textbf{Quantum isolator and circulator}\\
Quantum frequency conversion mediated by the beamsplitter-like Hamiltonian is fundamentally noiseless \cite{kumar1990quantum}. However, parasitic noise due to pump--matter interactions in the waveguide can still arise. We measure the on-chip noise to be linear in pump power, reaching approximately $2\times10^{-4}$ counts per second per Hz (cps/Hz) for both backward and forward directions at the signal wavelength, about 20 nm away from the pump, and at the pump power corresponding to maximum isolation (Fig. \ref{fig3}a). This noise level is well below the single-photon threshold of approximately 1 cps/Hz (one photon per time--frequency mode), confirming that our optical isolator can operate in the quantum regime.

To demonstrate the quantum-compatibility of our device, we first verify that signal isolation works for single photons. We use a pair of correlated non-degenerate photons in the telecom band with a bandwidth of 2.3 GHz generated from spontaneous parametric down-conversion (SPDC) and send the signal photon in the forward direction of the isolator. Fig. \ref{fig3}b shows the coincidence counts of the signal and idler photons when the pump is switched on and off. When the pump is off, the measured coincidence peak verifies the strong correlation between signal and idler; when the pump is on, the signal photons are blocked (by conversion to the 780-nm band and subsequent filtering), and consequently the coincidence vanishes. We then demonstrate that the time--energy entanglement of the SPDC photons is preserved after transmission through the optical isolator. As illustrated in Fig. \ref{fig3}c, we send a pair of degenerate SPDC photons with a bandwidth of 0.95~nm in the backward direction of the optical isolator and measure the time-resolved two-photon interference \cite{brendel1991time} via an unbalanced glass Mach--Zehnder interferometer (MZI). The MZI has a path delay of 1~ns, much longer than the coherence time of the SPDC photons, and its phase difference is controlled through temperature. The measured visibility of the two-photon interference fringe is 91.7\% limited by the imperfect interferometer visibility, which exceeds the Clauser--Horne limit of $1/\sqrt{2} \approx 70.7\%$ \cite{clauser1974experimental}, thereby confirming the time--energy entanglement of the transmitted photons (see the SI for measurement details).

\begin{figure}[!htb]
	\begin{center}
		\includegraphics[width=0.8\columnwidth]{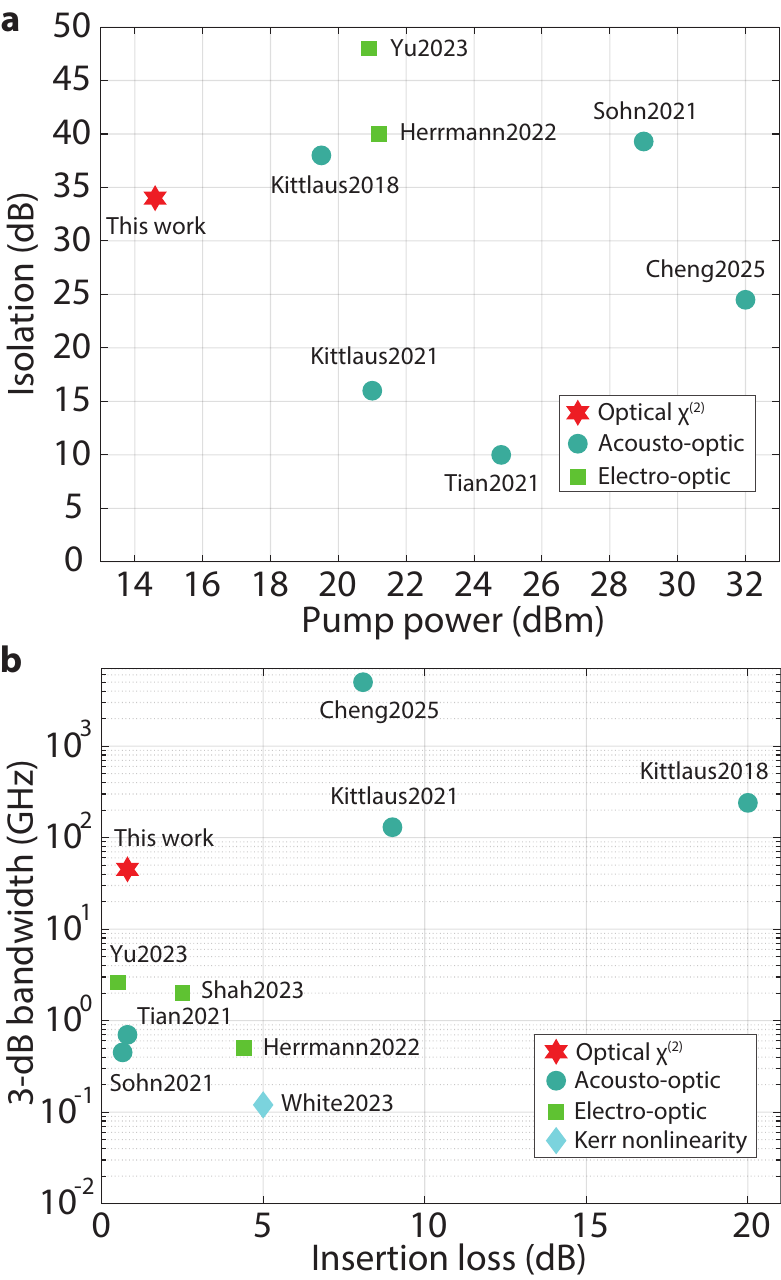}
		\caption{\textbf{Comparison of integrated nonmagnetic isolators.} 
			\textbf{a}. Isolation vs. pump power. \textbf{b}. 3-dB isolation bandwidth vs. insertion loss. Acousto-optic effect: \cite{cheng2025terahertz,tian2021magnetic,sohn2021electrically,kittlaus2021electrically,kittlaus2018non}. Electro-optic effect: \cite{shah2023visible,yu2023integrated,herrmann2022mirror}. Passive Kerr nonlinearity: \cite{white2023integrated}.  }
		\label{fig4}
	\end{center}
\end{figure}

We then utilize both modes $b$ and $c$ of the $\chi^{(2)}$ waveguide and the quantum frequency conversion between them to achieve optical circulation (see Fig. \ref{fig1}c). In this experiment, we use a 3.2-mm-long phase-matching-optimized waveguide. To characterize the optical circulation, we measure the transmission matrix $T_{ij}\equiv \langle \hat F_{i,\mathrm{out}}\rangle /\langle \hat F_{j,\mathrm{in}}\rangle$, where $\langle \hat F_{i,\mathrm{in(out)}}\rangle$ represents the input (output) photon flux of port $i$. Fig. \ref{fig3}d displays the measured transmission matrix for the four-port circulator with a 31-mW pump power (SI). To quantify the performance of the circulator, we calculate the operational fidelity \cite{scheucher2016quantum}
\be
\mathcal{F}=\frac{\mathrm{Tr}[\tilde T\cdot T_{\mathrm{id}}^T]}{\mathrm{Tr}[T_{\mathrm{id}}\cdot T_{\mathrm{id}}^T]},
\ee
where $\tilde T$ is the renormalized transmission matrix, $\tilde T_{ij}=T_{ij}/\sum_k T_{kj}$, and $T_{\mathrm{id}}$ is the transmission matrix of an ideal circulator. It can be shown that $0\leq\mathcal{F}\leq 1$ and for reciprocal transmission ($\tilde T=\tilde T^T$), the fidelity is bounded by $\mathcal{F}\leq0.5$ (SI). The fidelity of the measured circulator is $\mathcal{F}=0.962(4)$, while the average insertion loss is $\eta=-10\mathrm{log}\left(\frac{1}{4}\sum_{ij}T_{ij}\right)=2.8$ dB, which is limited by the 780-nm-band waveguide loss. The circulation direction can be reversed by changing the pump direction, and the measured transmission matrix for this case is shown in Fig. \ref{fig3}e with a fidelity of $\mathcal{F}=0.970(3)$. The added noise is $7.3\times 10^{-5}$ cps/Hz in the 1550-nm band and $2.5\times 10^{-5}$ cps/Hz in the 780-nm band via up-conversion, substantially lower than the single-photon level, indicating the circulator preserves quantum coherence and entanglement of the input photons. To explicitly demonstrate this, we use a pair of correlated non-degenerate photons in the telecom band with a bandwidth of 2.3 GHz and only send the signal photon in port 1. We measure the correlation between the up-converted signal in port 4 and the idler. The noise-subtracted coincidence counts ($C_{0}$) and total coincidence counts ($C_{t}$)  in the 2.3 GHz bandwidth are displayed in Fig. \ref{fig3}f. Based on this, the fidelity of the entangled photons after the circulator is found to be $F=\frac{1}{2}(1+\frac{C_{0}}{C_{t}})=0.968(6)$ (SI).

\noindent\textbf{Outlook}\\
In conclusion, we demonstrated a new form of optical nonreciprocity based on optical parametric frequency conversion that operates across a wide range of light levels---from individual photons to milliwatt powers---and spans a broad spectral range. Fig. \ref{fig4} compares key performance parameters of our isolator with other integrated nonmagnetic isolators demonstrated recently: our isolator achieves a remarkable isolation ratio with a low pump power while offering a unique combination of wide bandwidth and low insertion loss. Since III-V semiconductors have weak electro-optic and piezoelectric effects, our approach based on optical nonlinearity provides a solution for integrated isolators on III-V platforms. Moreover, the same strategy can be extended to other nonlinear photonic materials, such as periodically poled LiNbO$_3$, where extremely wide bandwidth of optical frequency conversion ($>1$ THz)---and thus isolation---can be achieved \cite{wang2023quantum}.

In terms of performance at the single-photon level, our integrated quantum optical isolator and circulator also significantly outperform previous demonstrations based on atomic systems and magneto-optic effects (see Tables \ref{tab:isolator} and \ref{tab:circulator}). Our approach, implemented directly on a scalable platform, enables the construction of robust classical and quantum optical circuits. For example, arranging multiple four-port circulators in one- or two-dimensional arrays can realize programmable networks that guide light in specific directions. Integration with quantum emitters, such as quantum-dot-embedded III-V semiconductors, would allow the construction of on-chip quantum light--matter interfaces with broken time--reversal symmetry \cite{gonzalez2015chiral,mahmoodian2020dynamics}. With its inherent reconfigurability and quantum frequency conversion capabilities, this platform offers new opportunities for directional quantum communication and noise-resilient quantum networks \cite{lodahl2017chiral,ahmadi2024nonreciprocal,almanakly2025deterministic}.

{\renewcommand{\arraystretch}{2.5}
\begin{table}[H] 
	\centering
	\caption{Quantum optical isolator performances}
	\label{tab:isolator}
	\resizebox{\columnwidth}{!}{\begin{tabular}{cccccc} 
		\hline
		\hline
		Scheme & Isolation & Bandwidth & Insertion loss  & Pump power & Reference \\ \hline
		\parbox{3.5cm}{Quantum frequency conversion}  & 34 dB  & 35.5 GHz  & 0.8 dB  & 27 mW  & This work \\
		Magneto-optic effect  & 25 dB  & 1.6 THz   & 20 dB  & /  & Ref. \cite{ren2022single} \\
		Single Rb atom  & 13 dB  & 230 MHz  & 1.4 dB    & /  & Ref. \cite{sayrin2015nanophotonic} \\
		Rb atomic gas   & 22.5 dB  & 200 MHz & 1.95 dB  &  100 mW & Ref. \cite{dong2021all} \\
		Rb atomic gas & 30.3 dB  &  175 MHz & 0.6 dB    &  40 mW  & Ref. \cite{zhang2023noiseless} \\ 
		\hline\hline
	\end{tabular}}
\end{table}

{\renewcommand{\arraystretch}{2.5}
\begin{table}[H] 
	\centering
	\caption{Quantum optical circulator performances}
	\label{tab:circulator}
	\resizebox{\columnwidth}{!}{\begin{tabular}{cccccc} 
		\hline
		\hline
		Scheme  & Fidelity & Average isolation & Bandwidth & Insertion loss & Reference \\ \hline
		\parbox{3.5cm}{Quantum frequency conversion} & 0.97  &  26 dB  & 48.8 GHz  & 2.8 dB   & This work \\
		Single Rb atom & 0.73   & 7 dB  & 0.23 GHz  & 1.4 dB  & Ref. \cite{scheucher2016quantum} \\
		\hline
		\hline
	\end{tabular}} 
\end{table}

\vspace{2mm}
\noindent\textbf{Acknowledgements}\\ 
This work is supported by US National Science Foundation under Grant No. 2223192 and QLCI-HQAN (Grant No. 2016136) and U.S. Department of Energy Office of Science National Quantum Information Science Research Centers. S. F. acknowledges the support of a MURI project from the Air Force Office of Scientific Research (Grant No. FA9550-22-1-0339).

\noindent\textbf{Author contributions}\\ 
K.F. conceived the experiment.  J.A., J.H. designed the device. J.A., J.H. fabricated the device.  J.H., H.Y., J.A. performed the experiment and analyzed the data.  All authors contributed to the writing of the paper.

%
 
\end{document}


\title{Supplementary Information for: Quantum-coherent optical isolation and circulation using frequency conversion on a chip}

\author{Jierui Hu} 
\thanks{These authors contributed equally to this work.}
\affiliation{Holonyak Micro and Nanotechnology Laboratory and Department of Electrical and Computer Engineering, University of Illinois at Urbana-Champaign, Urbana, IL 61801 USA}
\affiliation{Illinois Quantum Information Science and Technology Center, University of Illinois at Urbana-Champaign, Urbana, IL 61801 USA}
\author{Hao Yuan} 
\thanks{These authors contributed equally to this work.}
\affiliation{Holonyak Micro and Nanotechnology Laboratory and Department of Electrical and Computer Engineering, University of Illinois at Urbana-Champaign, Urbana, IL 61801 USA}
\affiliation{Illinois Quantum Information Science and Technology Center, University of Illinois at Urbana-Champaign, Urbana, IL 61801 USA}
\author{Joshua Akin} 
\thanks{These authors contributed equally to this work.}
\affiliation{Holonyak Micro and Nanotechnology Laboratory and Department of Electrical and Computer Engineering, University of Illinois at Urbana-Champaign, Urbana, IL 61801 USA}
\affiliation{Illinois Quantum Information Science and Technology Center, University of Illinois at Urbana-Champaign, Urbana, IL 61801 USA}
\author{Shanhui Fan} 
\affiliation{Department of Electrical Engineering, Ginzton Laboratory, Stanford University, Stanford, CA 94305, USA}
\author{Kejie Fang} 
\email{kfang3@illinois.edu}
\affiliation{Holonyak Micro and Nanotechnology Laboratory and Department of Electrical and Computer Engineering, University of Illinois at Urbana-Champaign, Urbana, IL 61801 USA}
\affiliation{Illinois Quantum Information Science and Technology Center, University of Illinois at Urbana-Champaign, Urbana, IL 61801 USA}

\maketitle

\tableofcontents

\newpage

\section{Device fabrication}

\begin{figure*}[!htb]
	\begin{center}
		\includegraphics[width=1\columnwidth]{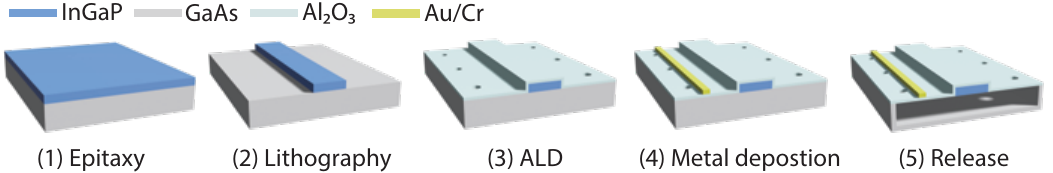}
		\caption{Illustration of the device fabrication process. }
		\label{fig::S1}
	\end{center}
\end{figure*}

The device fabrication process is illustrated in Fig. \ref{fig::S1}. The devices are fabricated from the 112 nm thick disordered In$_{0.5}$Ga$_{0.5}$P thin film grown on GaAs substrate by metal-organic chemical vapor deposition (T = 545 C, V/III = 48, precursors: trimethylindium, trimethylgallium and PH$_3$). The device pattern is defined using electron beam lithography and 150 nm thick negative tone resist hydrogen silsesquioxane (HSQ). A 20 nm thick layer of silicon dioxide is deposited on InGaP via plasma-enhanced chemical vapor deposition (PECVD) to promote the adhesion of HSQ. The device pattern is transferred to InGaP layer via inductively coupled plasma reactive-ion etch (ICP-RIE) using a mixture of Cl$_2$/CH$_4$/Ar gas. After a short buffered oxide etch to remove the residual oxide (both HSQ and PECVD oxide), a layer of 35 nm thick aluminum oxide is deposited on the chip via atomic layer deposition. A second electron beam lithography and subsequent ICP-RIE using CHF$_3$ gas are applied to pattern etch-through holes in the aluminum oxide layer for the undercut of the InGaP device. Next, a third electron beam lithography followed by electron-beam evaporation of 5 nm thick chromium and 20 nm thick gold is performed to define the electrodes. Finally, the InGaP device is released from the GaAs substrate using citric acid-based selective etching.

\section{Theoretical modeling of isolation}
Consider a waveguide supporting three optical modes, $a$, $b$, and $c$, satisfying $\omega_a + \omega_b = \omega_c$, where the $\chi^{(2)}$ nonlinearity facilitates a three-wave mixing process described by the Hamiltonian
$\hat H = \hbar \chi(\hat a^\dagger \hat b^\dagger \hat c + \hat a \hat b \hat c^\dagger)$
with $\hat a$ ($\hat a^\dagger$) denoting the annihilation (creation) operator for mode $a$, and similarly for modes $b$ and $c$. 
Suppose mode $a$ is driven by a classical pump field with amplitude $\alpha$, the interaction between modes $b$ and $c$ becomes linearized, i.e., $\hat H = \hbar \chi\alpha(\hat b^\dagger \hat c +  \hat b \hat c^\dagger)$.   In the forward direction, the coupled-mode equations describing the parametric frequency conversion between modes $b$ and $c$ is given by 
\begin{align}
\frac{\partial b}{\partial z}=-igc e^{-i\Delta k z}-\frac{\alpha}{2}b,\\
\frac{\partial c}{\partial z}=-igb e^{i\Delta k z}-\frac{\beta}{2}c,
\end{align}
where $g=\sqrt{\frac{\omega_b}{\omega_c}\eta_{\mathrm{SFG}}P_a}$, $\Delta k=k_c-k_a - k_b$, $\alpha$($\beta$) is the waveguide loss of mode $b(c)$, and $|b|^2(|c|^2$) is the photon flux of mode $b$($c$). The coupled-mode equations can be solved analytically. In the lossless case $\alpha=\beta=0$, the power in mode $b$ at length $L$ for $P_{c}(0)=0$ is given by 
\be\label{isolation}
\frac{P_b(L)}{P_b(0)}=\frac{\Delta k^2}{4g^2+\Delta k^2}+\frac{4g^2}{4g^2+\Delta k^2}\cos^2\sqrt{g^2+\frac{\Delta k^2}{4}}L.
\ee

In the backward direction, the modes do no satisfy the phase matching condition and thus no frequency conversion happens. The power of the input light is unchanged. As a result, the isolation ratio is given by Eq. \ref{isolation}.  When $g\gg \Delta k$, the maximum isolation ratio is found to be
\be
\mathcal{I}_{\mathrm{max}}=\left((2n+1)\frac{\pi}{L}\right)^2/\Delta k^2,
\ee
which is achieved when $\sqrt{g^2+\frac{\Delta k^2}{4}}=(n+\frac{1}{2})\frac{\pi}{L}$, for integer $n\geq 0$.

The fitting of the experimental data shown in the main text uses the solution of the coupled-mode equation for the general case with losses. 

\section{Phase-matching optimization}\label{sec:tuning}
\begin{figure*}[!htb]
	\begin{center}
		\includegraphics[width=0.8\columnwidth]{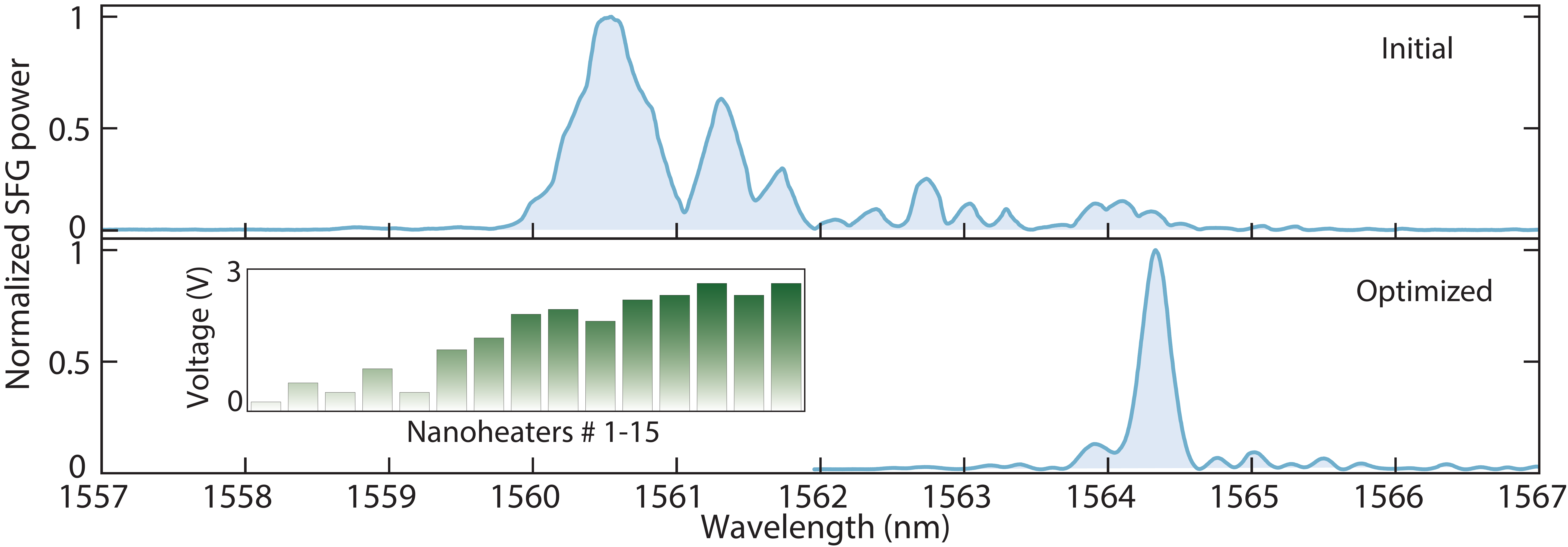}
		\caption{\textbf{Phase-matching optimization.} 
		        SFG spectrum of the 6-mm-long waveguide before and after optimization.
		        }
		\label{fig::S2}
	\end{center}
\end{figure*}

We use the nanoheater array to tune the phase-matching condition of the waveguide. Each nanoheater is 0.4-mm long, 400-nm wide, and 25-nm thick, which are individually controlled by a channel of a DC voltage source (NI 9264, 16-channel analog output module). Because of the large resistance of the nanoheater and the proximity to the waveguide, voltage of only up to a few volts is needed for each heater for the tuning. We monitor the sum-frequency generation (SFG) signal in the 780-nm-band while fixing the pump in the 1550-nm-band and scanning the wavelength of the 1550-nm-band signal.  
We begin by heating the first waveguide segment closest to the input, which separates the SFG signal generated in the first segment from the rest of the spectrum. Next, we heat the second segment and align its SFG spectrum with that of the first segment and maximize the combined SFG peak intensity. This procedure is then repeated sequentially for the remaining segments. In practice, a second round of fine tuning is usually performed to further optimize the aligned SFG signal from the first round. The SFG spectrum of the 6-mm waveguide before and after tuning is shown in Fig.~\ref{fig::S2}.
	
We then optimize the extinction of the input 1550-nm-band signal. We switch the detector to monitor the transmission in the 1550-nm-band and perform another round of phase-matching fine tuning. In this round, we focus on the segments close to the waveguide input to minimize the signal transmission, since they are less effective in tuning the SFG signal because of the large loss of 780-nm-band light but can tune the extinction of the 1550-nm-band signal effectively. 

\section{Self-calibrated frequency conversion efficiency}

To quantify the phase mismatch of a waveguide, one can calculate the self-calibrated efficiency ratio of frequency conversion processes, e.g., SFG and difference-frequency generation (DFG), defined as \cite{chen2024adapted,nash1970effect},
\begin{equation}\label{eq:RDefinition}
	R\equiv\frac{\eta}{\eta_0}=\frac{\eta\alpha}{AL},
\end{equation}
where $\eta$ and $\eta_0$ denote the peak efficiency of the measured and ideal frequency conversion spectrum, respectively, $A=\int \eta(\lambda)d\lambda$, and $\alpha=2\pi(\text{d}\Delta k/ \text{d}\lambda)^{-1}$ is the simulated dispersion factor with $\Delta k$ the wavevector mismatch between the modes for the respective conversion process. It should be noted that $\lambda$ in the definition of $A$ and $\alpha$ corresponds to the same mode. Because $R$ is related to the ratio of $\eta$ and $A$, it is free of calibration of the fiber coupling efficiency as well as the waveguide loss \cite{nash1970effect}. $R$ thus can be used to calibrate the quality of phase-matching tuning. We find $R=0.45$ for the 6-mm-long waveguide after optimization shown in Fig. \ref{fig::S2}. The discrepancy from the perfect phase matching condition is primarily due to the phase mismatch at length scales shorter than individual nanoheaters. In addition, the bandwidth of the ideal frequency conversion spectrum, denoted as $B_0$, follows the relation $B_0=\frac{5.57}{L}(\frac{\text{d}\Delta k}{\text{d}\Delta \lambda})^{-1}$ \cite{helmfrid1993influence}. The measured bandwidth $B$ is generally larger than $B_0$ due to the phase mismatch in the waveguide, and the ratio of $B_0/B$ can also be used to quantify the phase mismatch.

\section{Isolation of classical light}

\begin{figure*}[!htb]
	\begin{center}
		\includegraphics[width=1\columnwidth]{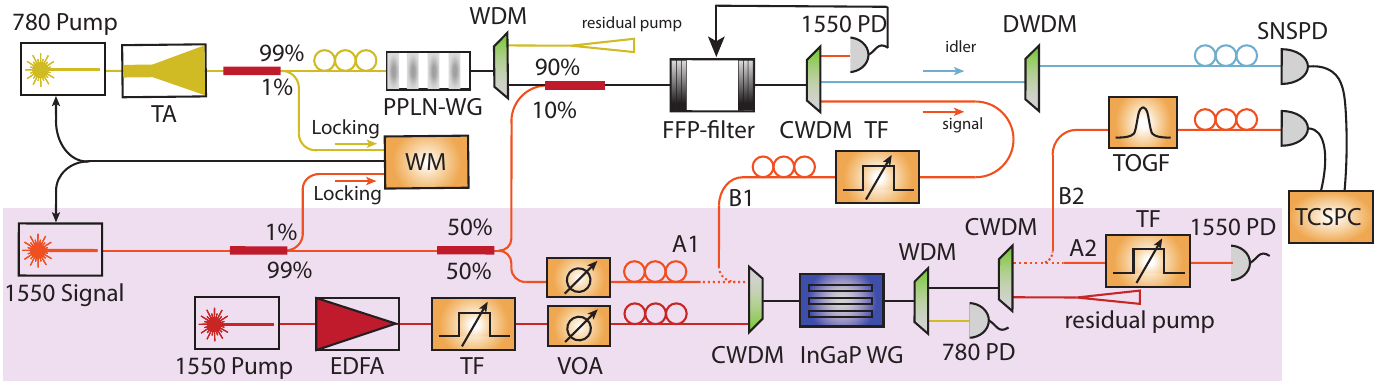}
		\caption{\textbf{Experimental setup for isolation measurement.} 
			InGaP WG: In$_{0.5}$Ga$_{0.5}$P waveguide. TA: tapered amplifier. PPLN-WG: periodically-poled lithium niobate waveguide. CWDM/DWDM: coarse/dense wavelength-division multiplexer. FFP filter: fiber Fabry-Perot filter. PD: photodetector. TF: tunable filter. WM: wavelength meter. EDFA: erbium-doped fiber amplifier. VOA: variable optical attenuator. A1/B1: signal input from laser/SPDC. A2/B2: signal output to PD/SNSPD  TOGF: tunable optical grating filter. SNSPD: superconducting nanowire single-photon detector. TCSPC: time-correlated single-photon counting module. The shaded area is for isolation of classical light only.}
		\label{fig::S3}
	\end{center}
\end{figure*} 

The experimental setup for isolation of classical light is shown in Fig.~\ref{fig::S3}. For the classical light isolation measurement, only the components within the shaded area are used. Ports A1 and A2 are connected to the coarse wavelength-division multiplexer (CWDM), while ports B1 and B2 are disconnected. The isolation pump is first amplified by an erbium-doped fiber amplifier (EDFA), followed by a tunable filter (TF) with a 1-nm bandwidth and a CWDM with a 20-nm bandwidth to suppress the amplified spontaneous emission noise. The filtered pump and signal are then combined using a CWDM and coupled into the device. At the output, the SFG signal is separated from the residual 1550-nm-band light using a WDM and directed to a 780-nm-band photodetector. The remaining 1550-nm-band signal is further separated from the 1550-nm-band pump by a CWDM and a 4-nm-bandwidth TF, and sent to a 1550-nm-band photodetector.

\begin{figure*}[!htb]
	\begin{center}
		\includegraphics[width=1\columnwidth]{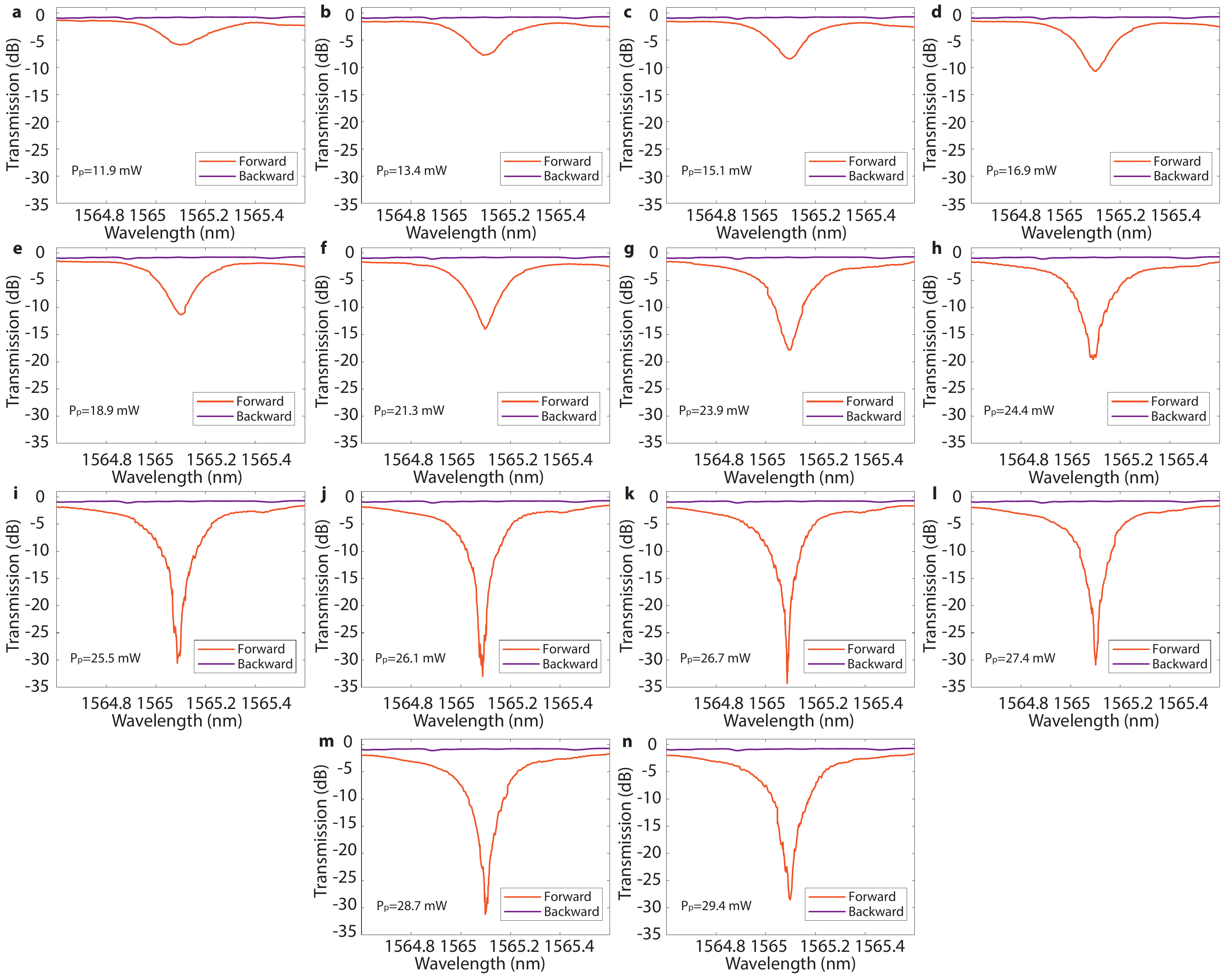}
		\caption{Forward and backward transmission corresponding to Fig. 2c for various pump power.
			}
		\label{fig::S4}
	\end{center}
\end{figure*}

Fig. \ref{fig::S4} shows the forward and backward transmission for various pump power corresponding to the peak isolation ratios in Fig. 2c.

\section{Isolation with different signal powers}

\begin{figure*}[!htb]
	\begin{center}
		\includegraphics[width=0.5\columnwidth]{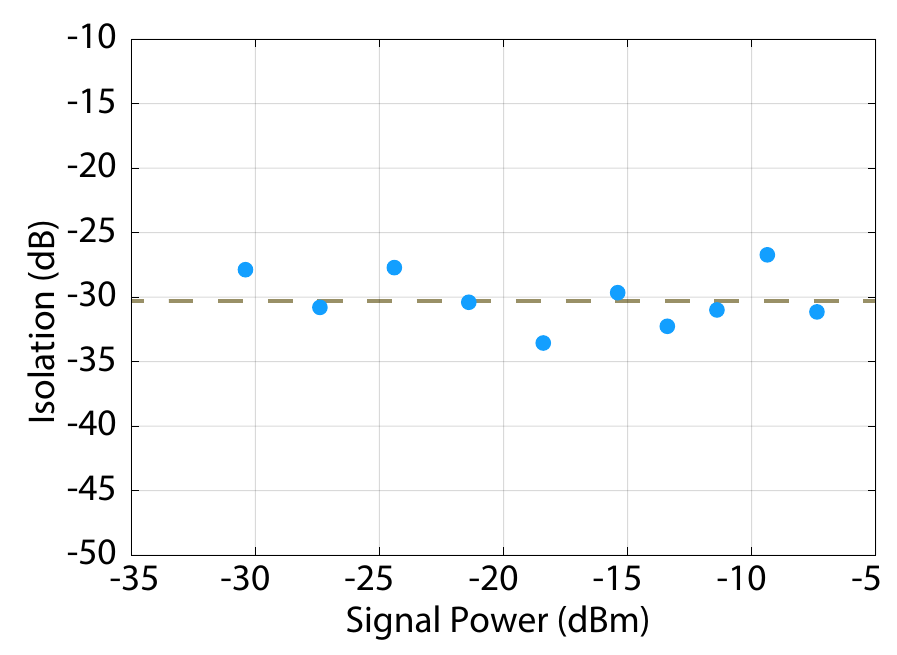}
		\caption{Isolation versus signal power. Dashed line is the average. }
		\label{fig::S5}
	\end{center}
\end{figure*}

We characterize the isolation performance of our device under different signal powers with the pump wavelength fixed at $\lambda_\text{p} = 1543$~nm. The result is shown in Fig. \ref{fig::S5}. The isolation is fairly stable with slight fluctuations as explained below. The phase-matching condition of the device is tuned at a signal power of $P_\text{signal} = 14.6~\mu$W. The signal power is then varied to measure isolation without a second tuning. Because high isolation is sensitive to the phase-matching condition, which is affected by the thermo-optic effect induced by the high-power pump, the pump power for each signal level is slightly adjusted to achieve maximum isolation, with an average pump power of 30~mW. In addition, the gain of the regular 1550-nm-band detector is adjusted to keep the signal within the acceptance range, which may also contribute to fluctuations in the measured isolation. An average isolation of 30.1~dB is obtained over a signal power range spanning 22~dB.

\section{On-chip noise} 

\begin{figure*}[!htb]
	\begin{center}
		\includegraphics[width=0.8\columnwidth]{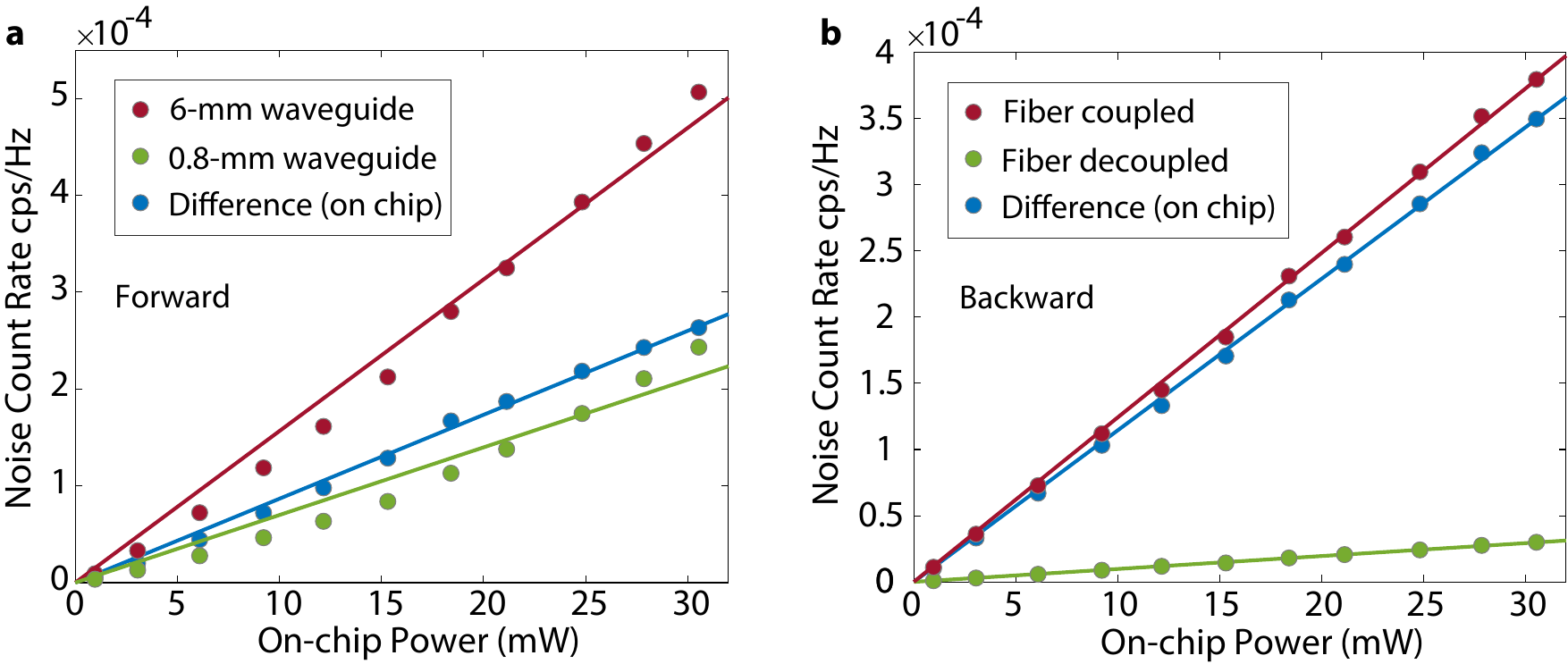}
		\caption{Measured noise for the forward and backward directions. }
		\label{fig::S6}
	\end{center}
\end{figure*}

To measure the parasitic noise associated with the pump, only pump is injected into the device. To characterize the noise in the backward direction, the reflected light from the device is filtered by a CWDM and a 4-nm-bandwidth TF to separate the residual pump, and subsequently detected using a SNSPD. The pump is set to a wavelength of 1543~nm, and its power is varied from 0 to 30~mW. Measurements are taken at detuning from the pump by approximately $\pm 20$ nm, and the count rates are found to be similar.  The on-chip noise is inferred through a differential method by subtracting the noise for the cases of fiber coupled and decoupled with the device, to remove noise that originates outside the device, e.g., in the fiber. Fig.~\ref{fig::S6}a shows the measured noise in the backward direction. To measure the noise in the forward direction, similar setup is used. However, in this case, to infer the on-chip noise, the differential method is subtraction of the noise of a long and a short waveguide, in order to remove the noise external to the chip. Fig.~\ref{fig::S6}b shows the measured noise in the forward direction.

Because the measured on-chip noise is linear in pump power and the pump frequency is below the bandgap, it could be attributed to spontaneous Raman scattering or defect-induced fluorescence. Though the two noise sources can be distinguished based on the fact that spontaneous Raman scattered light is polarized while fluorescence is not, our InGaP waveguide only support TE-modes in the 1550-nm band and consequently we cannot distinguish the two types of noises based on polarization---if both of them exist. However, since the InGaP used in our experiment is an undoped crystalline material, unlike amorphous materials, defect-induced florescence is expected to be low and the on-chip noise is likely dominated by spontaneous Raman scattering. 

\section{Isolation of single photons}

The experimental setup is shown in Fig. \ref{fig::S3}. For single-photon isolation, 
we tune the phase-matching condition of the device near the target isolation wavelength, following the procedure described in Section~\ref{sec:tuning}.  Then, ports A1 and A2 are disconnected, and ports B1 and B2 are connected to the CWDM, respectively. To prepare single photon, a 780-nm-band pump laser, amplified by a tapered amplifier (TA) and locked to 770.57~nm using a wavelength meter (WM), is used to generate spontaneous parametric down-conversion (SPDC) photon pairs using a periodically-poled lithium niobate (PPLN) waveguide. The residual 780-nm-band pump is filtered out using a WDM. The signal-idler photon pair is selected by a fiber Fabry-Perot (FFP) filter with a free spectral range (FSR) of 1147~GHz and a bandwidth of 2.3~GHz. 
The 1550-nm-band signal laser locked to the WM  
is used to stabilize the FFP. The locking laser is separated from the photon pair by a CWDM. Then, the signal and idler photons are further filtered by a 4-nm-bandwidth TF and a 0.94-nm-bandwidth dense
WDM (DWDM), respectively.

The idler photon is directed to a superconducting nanowire single-photon detector (SNSPD), while the signal photon is combined with the isolation pump and co-propagates through the device to realize isolation. A WDM
is applied after the device to separate the 780-nm-band SFG signal. A CWDM is then used to filter
the 1550-nm-band pump. A tunable optical grating filter (JDS TB9223, 3~dB bandwidth: 0.55~nm, 20~dB bandwidth: 1.5~nm) is applied to the signal photon path before detection to suppress broadband Raman noise generated in the device.  
To measure coincidence counts between the idler and signal photons, we use a time-correlated single-photon counting module (TCSPC), Swabian, to compute the correlation. For the pump-on case, we scan the isolation pump wavelength within a small range and collect coincidence data multiple times to identify the configuration yielding the highest isolation. The coincidence counts decrease from 10,599 to 273 when the pump is on, corresponding to an isolation of 16 dB, which is consistent with the isolation measured using a classical beam.

\begin{figure*}[!htb]
	\begin{center}
		\includegraphics[width=1\columnwidth]{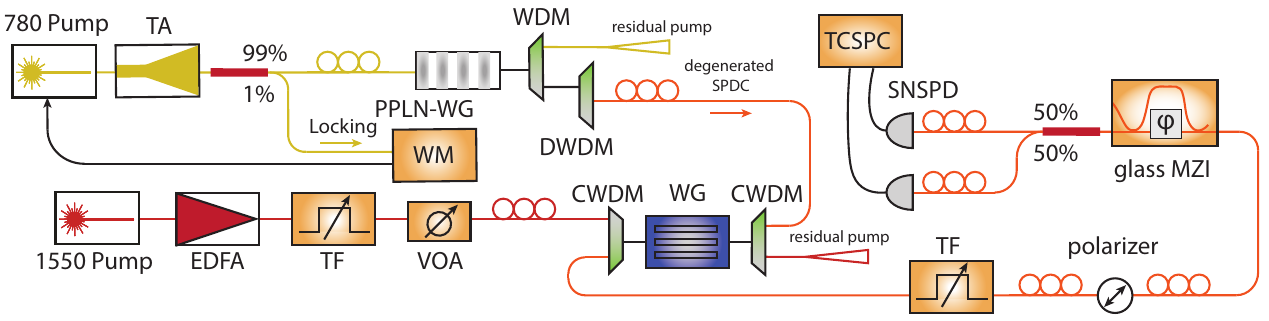}
		\caption{\textbf{Experimental setup for two-photon interference.} 
			WG: In$_{0.5}$Ga$_{0.5}$P waveguide. TA: tapered amplifier. PPLN-WG: periodically-poled lithium niobate waveguide. CWDM/DWDM: coarse/dense wavelength-division multiplexer. WM: wavelength meter. EDFA: erbium-doped fiber amplifier. TF: tunable filter. VOA: variable optical attenuator. MZI: Mach--Zehnder interferometer. SNSPD: superconducting nanowire single-photon detector. TCSPC: time-correlated single-photon counting module.}
		\label{fig::S7}
	\end{center}
\end{figure*}

The experimental setup for two-photon interference is shown in Fig.~\ref{fig::S7}. Before the experiment, a 1550-nm-band laser is used to optimize the visibility of the glass Mach--Zehnder interferometer (MZI), achieving a value of 97.4\%. A 780-nm-band laser is used to generate SPDC photon pairs. The 780-nm-band laser is 
locked by a wavelength meter, and then amplified by a TA before coupled into a PPLN waveguide. A combination of WDM and DWDM filters is used to eliminate the residual pump and select the degenerate SPDC photon pairs. The DWDM channel has a center wavelength of 1544.6~nm and a 3~dB bandwidth of 0.95~nm.

The isolation pump is 
amplified by an EDFA and coupled to the device with an on-chip power of approximately 30~mW. This pump counter-propagates with respect to the SPDC photon pair. After separating the incoming
1550-nm-band pump via a CWDM, a 4-nm-bandwidth tunable filter is used at the output of the device to suppress the broadband Raman noise. The photon pair then pass through a glass MZI, whose phase $\varphi$ is controlled by temperature. The path delay of the unbalanced MZI is $\tau_d = 1$~ns. The selected SPDC photons have a spectral bandwidth of 0.95~nm, corresponding to a single-photon coherence time much shorter than $\tau_d$. In contrast, the coherence time of the signal–idler photon pair, determined by the continuous-wave pump laser, is much longer than $\tau_d$. As a result, the photon pair can travel through either the short $(s)$ or long $(l)$ arm of the unbalanced interferometers, forming the entangled state $\left|\psi\right\rangle=\frac{1}{2}(\left|s\right\rangle_1 \left|s\right\rangle_2 +e^{i2\varphi} \left|l\right\rangle_1 \left|l\right\rangle_2)$, where $\varphi$ is the relative phase difference between the two paths for photons at the degenerate frequency. The entangled state is post-selected using time-resolved coincidence detection, with the central peak corresponding to a coincidence probability of $\tfrac{1}{4}\lvert 1+e^{i2\varphi}\rvert^2$. Finally, the photons are detected by SNSPDs and recorded with a TCSPC. By varying the MZI temperature, interference fringes in the coincidence counts are observed. The measured two-photon interference visibility is 91.7\%, limited by the visibility of the MZI of 97.4\%. Considering the MZI imperfections, the entangled-state visibility is expected to reach $90.4\%$~\cite{akin2024ingap}, which is close to our measured value.

For an unbalanced MZI consisting of two beam splitters, the transmission and reflection coefficients are denoted as $T_{1(2)}$ and $R_{1(2)}$, respectively. These coefficients satisfy the relation $T_{k}^2 + R_{k}^2 = 1$. The visibility of the interferometer, $V_1$, is limited by the deviation of the beam splitters from the ideal 50:50 ratio and can be measured using a continuous-wave laser. Similarly, the visibility of the two-photon interference fringe, $V_2$, is also limited by the imperfect values of $T_{1(2)}$ and $R_{1(2)}$, and is related to $V_1$ by \cite{akin2024ingap},
\begin{equation}\label{Visibility}
	V_2 = \frac{2}{\frac{4}{V_1^2} - 2}.
\end{equation}
We optimize the visibility of the MZI prior to the experiment and obtain $V_1 = 97.4\%$. Using Eq.~\ref{Visibility}, the theoretical two-photon interference fringe visibility can then be calculated as $V_2 = \frac{2}{4/V_1^2 - 2} = 90.4\%$.

\section{Optical circulation}

\begin{figure*}[!htb]
	\begin{center}
		\includegraphics[width=1\columnwidth]{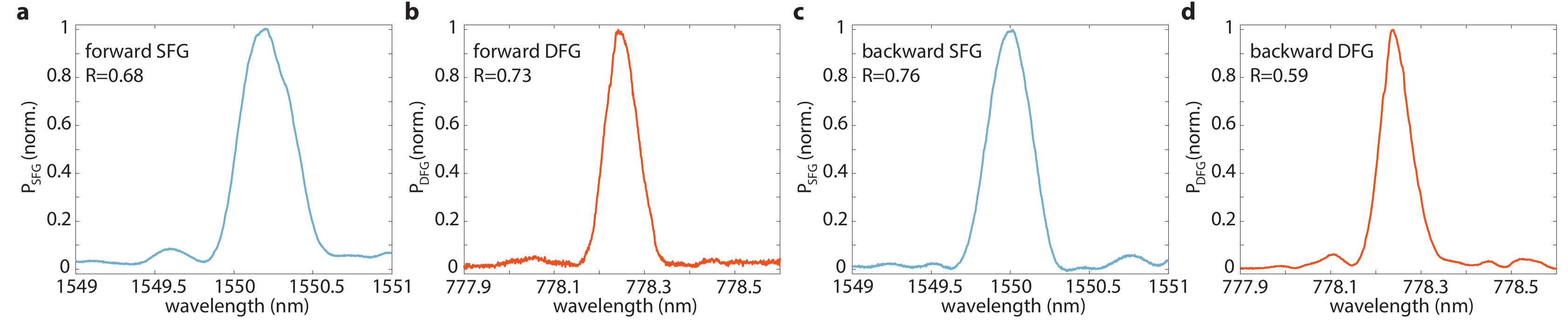}
		\caption{\textbf{Normalized frequency conversion spectrum.} \textbf{a.} Forward SFG spectrum. \textbf{b.} Forward DFG spectrum. \textbf{c.} Backward SFG spectrum. \textbf{d.} Backward DFG spectrum. The self-calibrated efficiency ratio $R$ is indicated.  }
		\label{fig::S8}
	\end{center}
\end{figure*}

For optical circulation, we use a 3.2-mm-long waveguide. We optimize the phase-matching condition of the forward SFG process using the nanoheater array. The resulting SFG spectrum is shown in Fig~\ref{fig::S8}a with a self-calibrated efficiency ratio of $R=0.68$. The forward DFG spectrum under the same condition is shown in Fig~\ref{fig::S8}b with a self-calibrated efficiency ratio of $R=0.73$. Using a 1563-nm pump with a power of 31~mW, we obtain a SFG conversion efficiency of 45.9\% and a DFG conversion efficiency of 46.3\%. In addition, we measure a 20~dB/cm loss in the 780-nm band and 1.3~dB/cm loss in the 1550-nm band. The operation bandwidth is 62 GHz.

To reverse the direction of optical circulation, we flip the direction of pump, without changing the nanoheater setting. The backward SFG and DFG spectra are shown in Figs.~\ref{fig::S8}c and d, respectively. The self-calibrated efficiency ratio of backward SFG and DFG is $R=0.76$ and $R=0.59$, respectively. The measured transmission matrices for the forward and backward pump are provided below,
\be
	T_{\mathrm{f}} =
	\begin{bmatrix}
		0              & 0.902\pm0.036 & 0              & 0      \\
		0.067\pm0.003  & 0             & 0.463\pm0.031  & 0      \\
		0              & 0             & 0              & 0.229\pm0.024 \\
		0.459\pm0.033  & 0             & 0.011\pm0.001  & 0
	\end{bmatrix},
\ee
\be
	T_{\mathrm{b}} =
	\begin{bmatrix}
		0              & 0.061\pm0.002  & 0             & 0.431\pm0.029  \\
		0.902\pm0.036 & 0              & 0             & 0      \\
		0              & 0.460\pm0.033  & 0             & 0.001\pm0.0001  \\
		0              & 0              & 0.229\pm0.024 & 0
	\end{bmatrix}.
\ee
The zero-entries of the transmission matrix are because there is no reflection and no frequency conversion opposite to the pump in the waveguide. The error is due to the fiber coupling efficiency uncertainty.

To quantify the performance of the circulator, we calculate the operational fidelity \cite{scheucher2016quantum}
\be
\mathcal{F}=\frac{\mathrm{Tr}[\tilde T\cdot T_{\mathrm{id}}^T]}{\mathrm{Tr}[T_{\mathrm{id}}\cdot T_{\mathrm{id}}^T]},
\ee
where $\tilde T$ is the renormalized transmission matrix, $\tilde T_{ij}=T_{ij}/\sum_k T_{kj}$, and $T_{\mathrm{id}}$ is the transmission matrix of an ideal circulator. The renormalization is necessary to remove the insertion loss, which can be calibrated separately via $\eta=-10\mathrm{log}\left(\frac{1}{4}\sum_{ij}T_{ij}\right)$. 

For counterclockwise circulation, the transmission matrix of the ideal circulator is 
\be
	T_{\mathrm{id}} =
	\begin{bmatrix}
		0              & 1 & 0              & 0      \\
		0  & 0             & 1 & 0      \\
		0              & 0             & 0              & 1 \\
		1  & 0             & 0 & 0
	\end{bmatrix}.
\ee
Thus, $\mathcal{F}=\frac{1}{4}(\tilde T_{12}+\tilde T_{23}+\tilde T_{34}+\tilde T_{41})\leq \frac{1}{4}\sum_{ij}\tilde T_{ij}=1$, where the upper bound 1 is saturated for ideal circulation. Furthermore, for reciprocal transmission ($\tilde T=\tilde T^T$), $\mathcal{F}=\frac{1}{8}(\tilde T_{12}+\tilde T_{21}+\tilde T_{23}+\tilde T_{32}+\tilde T_{34}+\tilde T_{43}+\tilde T_{41}+\tilde T_{14})\leq \frac{1}{8}\sum_{ij}\tilde T_{ij}=\frac{1}{2}$. For the measured circulation with forward and backward pump, the fidelity is $\mathcal{F}=0.962(4)$ and $\mathcal{F}=0.970(3)$, respectively. 

The on-chip noise is measured to be $n_{15}=7.3\times10^{-5}$ cps/Hz in the 1550-nm band under the operation condition of the circulator. The up-converted noise in the 780-nm band can be calculated using the frequency conversion efficiency including the waveguide loss, and is found to be $n_{78}=2.5\times10^{-5}$ cps/Hz. The high conversion efficiency and low added noise allow the optical circulator to be operated in the quantum regime. We verify that the SFG process preserves the time--energy entanglement of a pair of SPDC photons when the signal photon is up-converted to the 780-nm band (see Section \ref{QFCcorrelation}). We also verify that the DFG process preserves single-photon coherence by performing quantum state tomography of a down-converted time-bin 780-nm qubit generated from attenuated laser pulses \cite{hu2025high}. 

The average isolation of our circulator is calculated using 
\be
\frac{1}{4}\big(10\mathrm{log}(T_{\mathrm{f},12}/T_{\mathrm{f},21})+10\mathrm{log}(T_{\mathrm{f},34}/T_{\mathrm{f},43})+10\mathrm{log}(T_{\mathrm{f},23}/n_{78})+10\mathrm{log}(T_{\mathrm{f},41}/n_{15})\big).
\ee
Since there is no transmission for $2\rightarrow 3$ and $4\rightarrow 1$ for the counterclockwise circulation, we use the added noise in the 1550-nm and 780-nm bands to replace the transmission for these two directions while the input signal is assumed to be single photons, i.e., 1 cps/Hz.

\section{Photon correlations via frequency conversion}\label{QFCcorrelation}

\begin{figure*}[!htb]
	\begin{center}
		\includegraphics[width=1\columnwidth]{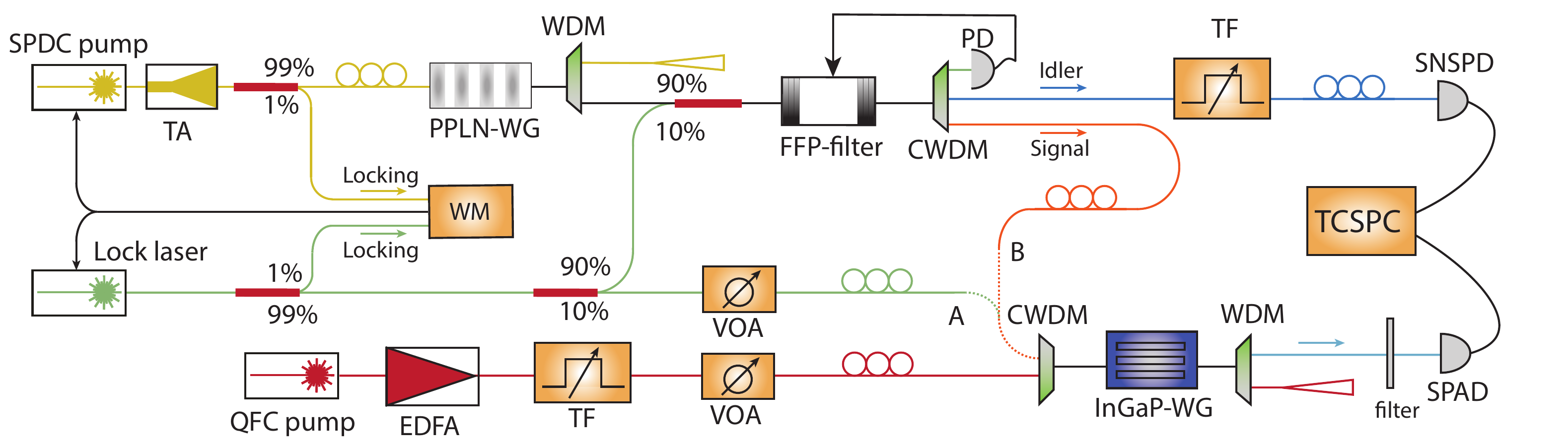}
		\caption{\textbf{Experimental setup for Photon correlations via QFC.} 
			WG: In$_{0.5}$Ga$_{0.5}$P waveguide. TA: tapered amplifier. PPLN-WG: periodically-poled lithium niobate waveguide. CWDM: coarse wavelength-division multiplexer. WM: wavelength meter. EDFA: erbium-doped fiber amplifier. TF: tunable filter. VOA: variable optical attenuator. SNSPD: superconducting nanowire single-photon detector. SPAD: single-photon avalanche diode detector. TCSPC: time-correlated single-photon counting module.}
		\label{fig::S9}
	\end{center}
\end{figure*}

The experimental setup for the photon correlation measurement via QFC of the circulator is shown in Fig.~\ref{fig::S9}. A 780-nm-band laser is amplified by a TA and serves as the pump source for SPDC. At the output of the PPLN waveguide, three 780/1550-nm WDMs are used to suppress residual 780-nm pump light. The signal-idler photon pairs are filtered by a fiber FFP filter with a FSR of 1147~GHz and a bandwidth of 2.3~GHz. The wavelengths of the idler and signal photons are 1567.64~nm and 1521.31~nm, respectively. A 1550-nm-band laser, locked to the WM, is used to stabilize the FFP. The locking laser is separated from the photon pairs by a CWDM. The idler photons are further filtered by a 4-nm-bandwidth TF and detected using a SNSPD. The signal photons are sent into the InGaP waveguide, upconverted to the 780-nm band, and detected by a SPAD. Finally, the correlations between the photons are recorded using a TCSPC.

The phase-matching condition of the 3.2-mm-long waveguide is tuned prior to the correlation measurement. The lock laser is used as the signal laser to optimize the SFG spectrum, with port~A connected to the CWDM before the device and port~B left unconnected. After tuning the device, the QFC pump wavelength is finely adjusted while monitoring the SFG signal and reading the wavelength from the WM, ensuring that the SFG efficiency is optimized at the photon-pair signal wavelength.

To suppress noise during the measurement, a 1-nm-bandwidth TF and a CWDM are used to filter out the ASE from the EDFA before combining the pump with the signal path. Additionally, three filters are employed before the SPAD to remove the SHG component of the pump. We then measure the photon correlations between the idler and the converted signal photons under different SPDC rates, with port~B connected and port~A disconnected.

\begin{figure*}[!htb]
	\begin{center}
		\includegraphics[width=0.4\columnwidth]{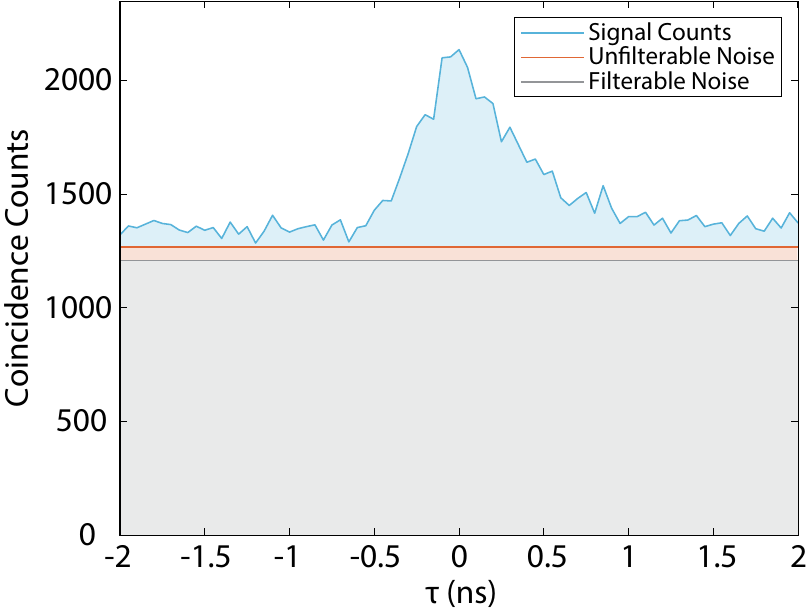}
		\caption{\textbf{Collected coincidences for photon correlations via QFC.} 
			Directly measured coincidence counts for the highest SPDC rate in the main text of 144 MHz for 120 s. Coincidence counts can be separated into three distinct sources: signal coincidence (blue), noise coincidences in the signal bandwidth (orange), and noise coincidences outside the signal bandwidth (gray).}
		\label{fig::S10}
	\end{center}
\end{figure*}

The coincidence counts are processed from the raw coincidences collected by the TCSPC at the end of the setup. A raw collected coincidence plot is shown in Fig.~\ref{fig::S10}. The collected coincidences can be separated into three sources: 1. signal coincidences from upconverted SPDC photons (blue), 2. upconverted noise coincidences residing in the same bandwidth as the signal (orange), and 3. upconverted noise coincidences that have distinctly different frequencies than the signal (gray). The last part can be effectively removed by proper filters. Figure 4f in the main text only includes noise that is unfilterable and therefore fundamental to the device.

The signal coincidences should follow the same second order correlation relation as the initial SPDC source with an accompanying background of accidental coincidence counts. This raw correlation is measured first to analyze the input coincidences. The noise can be measured independently by disconnecting port B and only allowing the QFC pump to propagate in the device. With no signal the detected counts are upconverted broadband Raman noise. The device QFC bandwidth of 49 GHz defines the total noise that is upconverted. Most of this noise in theory can be filtered and the only fundamental noise added to the signal is the upconverted noise residing in the same 2.3 GHz bandwidth as the SPDC signal. The coincidences referenced in the main text are specifically taken at zero time delay. To process the data into the figure in the main text first the noise is measured independently and subtracted to generate the noise-subtracted data. Then the total coincidence data adds back the ratio of noise inside the signal bandwidth that is fundamental to the QFC process. All data has  losses from the device to the detector factored out to represent the coincidences just at the end of the QFC device.

\section{Fidelity of time-energy entangled photons}
The measured fidelity of the time-energy entangled photon pair can be inferred from the visibility of the two-photon interference using Franson interferometer. Suppose only the signal is subject to QFC, as in our experiment, then the coincidence between signal and idler is given by
\be
C[\varphi]\propto\frac{1}{4}g^{(2)}(0)R_sR_i|1+e^{2i\varphi}|^2+\frac{1}{2}R_iR_n,
\ee
where $R_s$ and $R_n$ are the rates of signal and noise on chip in the same bandwidth $B$, $g^{(2)}(0)$ is the zero-delay normalized correlation function between signal and idler, and $\varphi$ is the interferometer phase.

The visibility of the two-photon fringe is given by
\be
V=\frac{C[0]-C[\pi]}{C[0]+C[\pi]}=\frac{g^{(2)}(0)R_sR_i}{g^{(2)}(0)R_sR_i+R_iR_n}=\frac{C_0}{C_0+C_{U}}=\frac{C_0}{C_t},
\ee
where $C_0=g^{(2)}(0)R_sR_i\Delta t T$ is the desired coincidence counts,  $C_{U}=R_nR_i\Delta t T$ is the undesired coincidence counts, and $C_t=C_0+C_U$ is the total coincidence counts. $\Delta t$ is the coincidence time-bin width and $T$ is the integration time. 
The fidelity of the entangled-photon pair is inferred as \cite{tanzilli2005photonic}
\be\label{Fdef1}
F=\frac{1+V}{2}=\frac{1}{2}\left(1+\frac{C_0}{C_t}\right).
\ee
We used this equation to calculate the fidelity of the entangled photons after the circulator, as described in the previous section.

Using $g^{(2)}(0)=\frac{B}{R_{e0}}=\frac{B}{R_s/\eta}$, where $R_{e0}$ is the entangled-pair generation rate in the bandwidth $B$ at the source and $\eta$ is the transmission efficiency from the entangled-photon source to the input of the QFC device, Eq. \ref{Fdef1} can be expressed as  
\be
F=\frac{2+\frac{R_n}{\eta B}}{2(1+\frac{R_n}{\eta B})}.
\ee
The measured fidelity thus is fundamentally limited by the noise rate per bandwidth $\frac{R_n}{B}$ and is independent of the entangled photon rate. In our experiment, $\frac{R_n}{B}\approx 10^{-4}$ (before QFC) and $\eta\approx 0.03$, so $F>0.99$ theoretically. We also point out the measured fidelity can be increased by improving transmission efficiency $\eta$ of the experiment setup, which is not a fundamental limit of the device itself.